\documentclass{aa}
\usepackage{graphicx}
\pdfoutput=1
\usepackage[varg]{txfonts}  
\usepackage[section]{placeins}
\usepackage{subcaption}
\usepackage{booktabs}
\usepackage[utf8]{inputenc}
\titlerunning{An adaptive parameter estimator for poor-quality spectral data of WDs}
\authorrunning{Duo Xie et al.}
\usepackage[colorlinks=true, linkcolor=blue, urlcolor=blue, citecolor=blue]{hyperref}
\begin{document}

\title{An adaptive parameter estimator for poor-quality spectral data of white dwarfs
}

\author{Duo Xie\inst{1}
  \and Jiangchuan Zhang\inst{2}\thanks{These authors contributed equally to this work.}
  \and Yude Bu\inst{2}\thanks{Corresponding author.}
  \and Zhenping Yi\inst{3}
  \and Meng Liu\inst{3}
  \and Xiaoming Kong\inst{3}
}

\institute{
  School of Business, Shandong University, Weihai, 264209, Shandong, People’s Republic of China
  \and
  School of Mathematics and Statistics, Shandong University, Weihai, 264209, Shandong, People’s Republic of China
  \email{buyude@sdu.edu.cn}
  \and
  School of Mechanical, Electrical and Information Engineering, Shandong University, Weihai, 264209, Shandong, People’s Republic of China
}

\abstract{White dwarfs represent the end stage for 97\% of stars, making precise parameter measurement crucial for understanding stellar evolution. Traditional estimation methods involve fitting spectra or photometry, which require high-quality data. In recent years, machine learning has played a crucial role in processing spectral data due to its speed, automation, and accuracy. However, two common issues have been identified. First, most studies rely on data with high signal-to-noise ratios (SNR > 10), leaving many poor-quality datasets underutilized. Second, existing machine learning models, primarily based on convolutional networks, recurrent networks, and their variants, cannot simultaneously capture both the spatial and sequential information of spectra. To address these challenges, we designed the Estimator Network (EstNet), an advanced algorithm integrating multiple techniques, including Residual Networks, Squeeze and Excitation Attention, Gated Recurrent Units, Adaptive Loss, and Monte-Carlo Dropout Layers. We conducted parameter estimation on 5,965 poor-quality white dwarf spectra (R~1800, SNR~1.17), achieving average percentage errors of 14.86\% for effective temperature and 3.97\% for surface gravity. These results are significantly superior to other mainstream algorithms and consistent with the outcomes of traditional theoretical spectrum fitting methods. In the future, our algorithms will be applied for large-scale parameter estimation on the Chinese Space Station Telescope and the Large Synoptic Survey Telescope.}


\maketitle

\section{\textbf{Introduction}}

Accurately measuring the atmospheric physical parameters of white dwarfs is crucial for studying stellar evolution and the structure of the Milky Way, given the significant roles white dwarfs play in various astronomical phenomena. For instance, Type Ia supernovae, which result from the explosion of white dwarfs in close binary systems, are key to understanding cosmic distances \citep{qi2022past, bildsten2007faint}. Double white dwarfs are the primary sources of gravitational waves detected by the Laser Interferometer Space Antenna (LISA) \citep{finch2023identifying}. Additionally, the luminosity function of white dwarfs is used to estimate the ages of star clusters \citep{bedin2019hst, cukanovaite2023local}, and their mass function can measure the stellar death rate and estimate the age of the galactic halo \citep{liebert2005formation, kalirai2013applications}. Proper motion data of white dwarfs aids in studying the distribution of dark matter in the galactic halo, providing crucial insights into its properties \citep{torres2002high}. Thus, white dwarfs are vital for investigating stellar evolution and the structure of the Milky Way.

   \begin{figure*}
   \centering
   \includegraphics[width=\hsize]{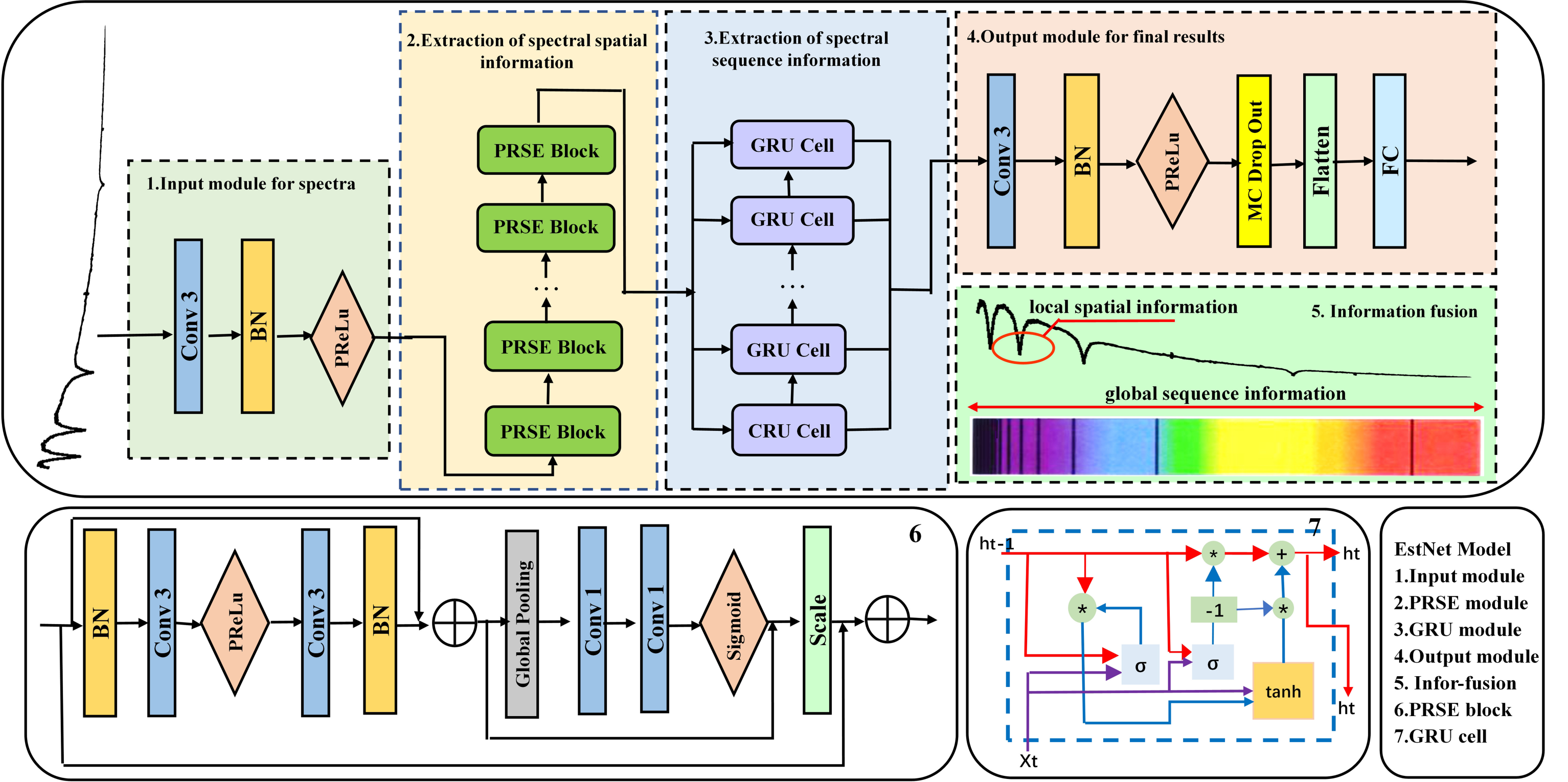}
      \caption{EstNet Architecture. EstNet comprises four parts: Input Module, PRSE Module, GRU Module, and Output Module, as shown in parts 1 to 4. Part 5 illustrates two types of information present in spectra: the local spatial information of a specific spectral line and the global dependencies between different absorption lines. Part 6 describes the structure of the PRSE block, which determines the depth of EstNet. Part 7 details the internal structure of the GRU cell, a neural component with memory capabilities.}
         \label{fig:EstNet}
   \end{figure*}

White dwarfs are compact stars with low luminosity, high density, high surface temperature, blue-white color, and radiation energy concentrated in the ultraviolet band. Approximately 97\% of stars in the Milky Way will eventually evolve into white dwarfs \citep{parsons2020pulsating, almeida2023eighteenth}. The initial mass of white dwarfs is about $8.5M_{\odot}$ to $10.6M_{\odot}$, and they experience mass loss during evolution. Currently, observed white dwarfs have masses primarily ranging from $0.5M_{\odot}$ to $0.8M_{\odot}$ \citep{woosley2015remarkable}. Additionally, \citet{wang2022extremely} discovered 21 white dwarfs with masses not exceeding $0.3M_{\odot}$ in the Gaia DR2 and LAMOST DR8 data. The effective temperature of white dwarfs ranges from 4000K to 150000K, with an average surface gravity acceleration of $10^8 \text{cm s}^{-2}$ (log g = 8), and their apparent magnitudes are concentrated at $15.5^m$. The faintest known white dwarf has a luminosity of $10^{-4.7}L_{\odot}$ \citep{fontaine2001potential}, and most have luminosities between $10^{-2}L_{\odot}$ and $10^{-3}L_{\odot}$ \citep{mccook1999catalog}. Their radii typically range from 0.008 to 0.023$R_{\odot}$ \citep{fontaine2001potential}. This paper focuses on estimating surface gravity \(\log\) g and effective temperature \(T_{\text{eff}}\).

Traditional methods generally have higher estimation accuracy for stellar atmospheric physical parameters but are less efficient when dealing with large amounts of poor-quality data. For photometric data, traditional methods use color indices to calculate atmospheric physical parameters for different spectral types of stars \citep{crawford1958two, alonso1999effective, kirby2008metallicity}. Color indices are calculated based on photometric magnitudes in various bands, which can be influenced by interstellar reddening and extinction, thereby affecting the calculation. For spectral data, atmospheric physical parameters are mostly estimated through template matching \citep{lee2011segue, barbuy2003grid}, or line index methods \citep{thomas2002epochs, ting2019payne}. Template matching relies on the distance between theoretical and observed spectra and requires high-quality observed spectra. Low-resolution, low signal-to-noise ratio spectra contain a lot of noise. The line index method relies on flux information at specific wavelengths and does not consider the entire spectrum, limiting its effectiveness when spectral data is incomplete.

Recently, there have been advancements in estimating the parameters of white dwarfs. \citet{chen2024detecting} leveraged Bayesian methods to estimate the parameters of white dwarfs. \citet{liang2024revisiting} estimated the stochastic gravitational-wave background (SGWB) of white dwarfs using weak-signal limits. \citet{suleimanov2024application} employed hydrostatic LTE atmosphere models to estimate the mass of white dwarfs. \citet{castro2024fast} developed a mixing length theory (MLT) incorporating thermal diffusion and composition gradients, analyzing crystallization-driven convection in carbon-oxygen white dwarfs. \citet{ferreira2024benchmark} estimated the physical and dynamical parameters of white dwarfs through interpolations with theoretical models and evolutionary tracks. \citet{panthi2024uocs} used TESS data to construct light curves and, combined with multi-wavelength spectral energy distributions (SEDs), estimated the luminosity, temperature, and radius of white dwarfs. Theoretical models are established under certain assumptions, such as isolated stars or stars in binary systems with uniformly distributed ambient media, no mass loss, and no complex material exchange. However, these assumptions may not fully match actual observations, and recent work has focused on further optimizing white dwarf atmospheric models \citep{chen2022asteroseismology, ferrand2022double, camisassa2022evolution}.

\begin{figure*}
    \centering
    \begin{subfigure}{0.33\textwidth}
        \centering
        \includegraphics[width=\textwidth]{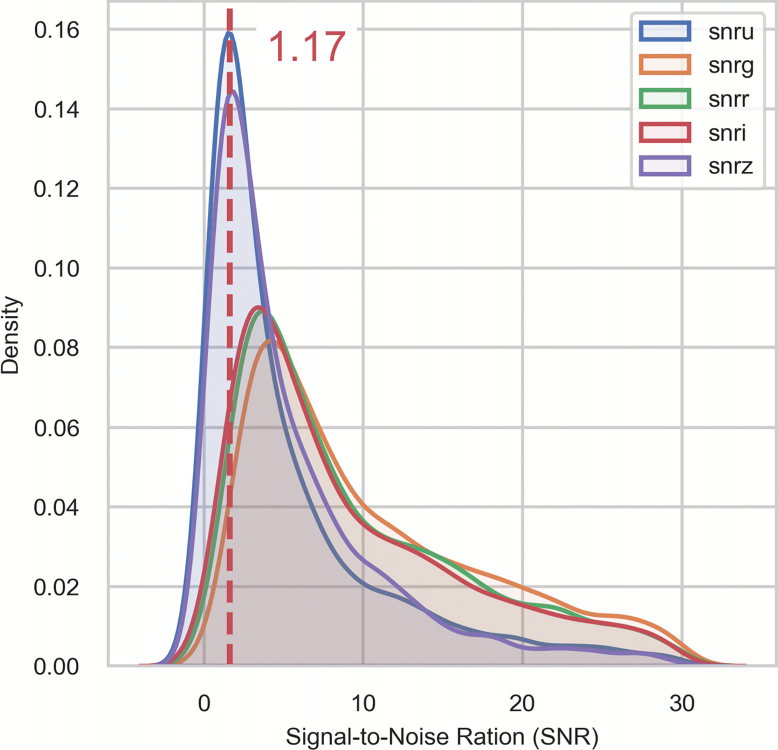}
        \caption{}
        \label{fig:fig1}
    \end{subfigure}%
    \begin{subfigure}{0.33\textwidth}
        \centering
        \includegraphics[width=\textwidth]{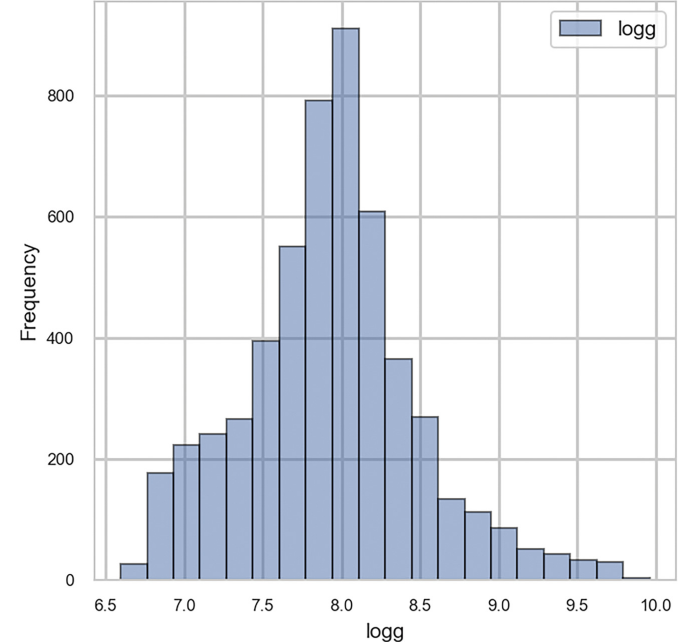}
        \caption{}
        \label{fig:fig2}
    \end{subfigure}%
    \begin{subfigure}{0.33\textwidth}
        \centering
        \includegraphics[width=\textwidth]{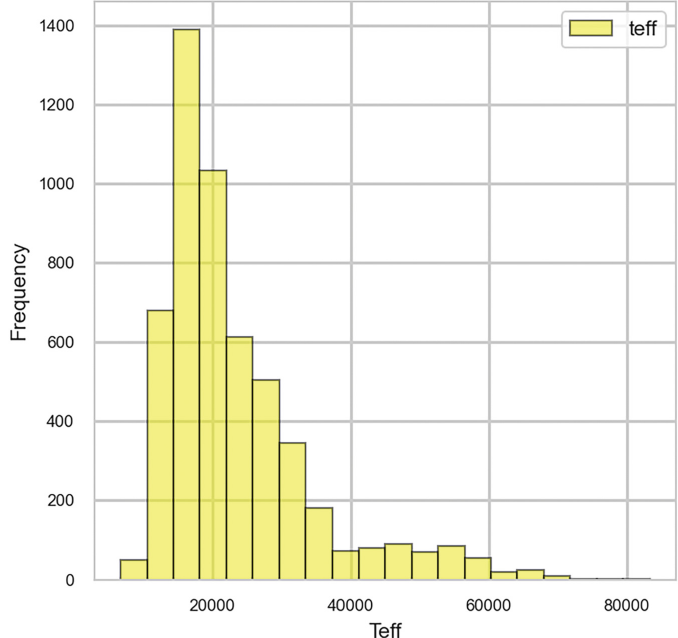}
        \caption{}
        \label{fig:fig3}
    \end{subfigure}
    \caption{Data Description. (a) shows the signal-to-noise ratio (SNR) distribution across the \(u, g, r, i, z\) bands from LAMOST. White dwarf characteristics are primarily concentrated in the \(u\) band, with a median SNR of 3.14 and a mode of 1.17, indicating extremely poor data quality. (b) shows the distribution of \(\log\) g labels, primarily concentrated around 8 dex. (c) shows the distribution of \(T_{\text{eff}}\) labels, primarily concentrated around 20000 K.}
    \label{fig:fig2}
\end{figure*}

With the continuous release of large-scale survey data, the volume of astronomical data is expected to grow exponentially. The vast amount of spectral data requires rapid, efficient, and timely processing. Traditional astronomical methods, which rely on expert knowledge and manual supervision to estimate atmospheric physical parameters, will face significant challenges. However, the rapid development of deep learning technology offers new opportunities for estimating these parameters. \citet{smith2023astronomia} argued that the symbiotic relationship between large, high-quality, multi-modal public datasets in astronomy and cutting-edge deep learning research can mutually promote each other. Neural networks, a crucial component of deep learning, simulate the neural circuits of human brain and mathematically characterize the Hebbian Rule \citep{sejnowski1989hebb}. With the introduction of backpropagation algorithms \citep{lecun1988theoretical}, solutions to the vanishing gradient problem \citep{hochreiter1998vanishing, hu2018overcoming}, and improvements in computer hardware resources, neural networks have become indispensable in AI \citep{kumar2012advanced, wu2018development}, impacting astronomical research as well.

In recent years, many researchers have used deep learning techniques to estimate stellar parameters, such as fully connected neural network (FCN) \citep{chandra2020computational, leung2019simultaneous}, convolutional neural network (CNN) \citep{minsky1961steps, wu2024estimating}, and recurrent neural network (RNN) \citep{li2023estimating, flores2021stellar}. However, each of these deep learning networks has its limitations. FCN has a large number of parameters, requiring high computational and storage resources, making them unsuitable for feature extraction as the backbone network \citep{alqahtani2021literature}. CNN focuses on the local spatial information of input data in each layer and perform poorly when processing data with sequential dependencies \citep{zuo2015convolutional}. RNN can handle and capture the time step sequence information of input data, but they may encounter gradient vanishing and exploding problems when capturing long-term dependencies \citep{hochreiter1997long}.

The spectra of white dwarfs contain many characteristic absorption lines, with DA-type white dwarfs exhibiting prominent Balmer absorption lines \citep{mccook1987catalog}. Each type of spectral line is distributed at different wavelengths and is interrelated. In manual spectral classification, researchers analyze these lines to classify the spectral types of white dwarfs or estimate their parameters by fitting the Balmer absorption lines \citep{zhao201370}. For a white dwarf spectrum, It is essential to locate the local spatial information of specific absorption lines, while taking into account the dependencies among them, commonly known as sequence information. Therefore, there is an urgent need for an algorithm that can extract both spatial and sequence information from spectra.

High-quality data enables researchers to provide more accurate parameter estimates, but such data is often scarce within the vast amounts of survey data. Therefore, extracting valuable information from poor-quality data is crucial. Raw observed data must undergo a series of processes, including extraction, calibration, and sky subtraction, to generate one-dimensional spectra. Failing to utilize poor-quality data results in significant data waste. Currently, most parameter estimation tasks rely on high signal-to-noise ratio (SN > 10) spectral data \citep{flores2021stellar, guo2022white, xiang2022stellar, kong2017three, tremblay2011improved}. The characteristic absorption lines of white dwarfs are mainly concentrated in the blue end (3900Å-5900Å) \citep{kong2018spectral}. In our study, the spectral data has a resolution of 1800, with a median signal-to-noise ratio of 3.135 and a mode of 1.17 in the u-band, classifying it as poor-quality spectral data. Low-resolution, low signal-to-noise ratio data exhibits poorer quality, with missing flux and significant noise. To address this issue, we propose an adaptive loss mechanism that enables the model to focus more on learning from non-anomalous data and less from anomalous data, thereby improving the robustness of parameter estimation.

The main work of this paper is developing the Estimator Network (EstNet) for estimating the atmospheric physical parameters of white dwarfs surface gravity \(\log\) g and effective temperature \(T_{\text{eff}}\). The four main contributions are as follows:
\begin{enumerate}
    \item We proposed an adaptive loss mechanism that drives EstNet to learn more from non-anomalous data, reducing its reliance on anomalous data and decreasing the model's sensitivity to anomalies.
    \item EstNet integrates the design principles of CNN and RNN, enabling it to capture both local spatial information and global sequence information from spectral data. This approach overcomes the limitations of existing methods and enhances the model's ability to learn from poor-quality data.
    \item The output of EstNet is no longer a single point estimate but an interval estimate of the labels, allowing for the measurement of model uncertainty and improving interpretability.
    \item We conducted reliability analysis, comparative analysis, robustness analysis, and validation analysis on EstNet, demonstrating its reliability and effectiveness from multiple perspectives.
\end{enumerate}

The structure of this paper is as follows: Section \hyperref[sec:Methods]{\ref*{sec:Methods}} introduces the EstNet model framework. Section \hyperref[sec:Data]{\ref*{sec:Data}} describes the datasets. Section \hyperref[sec:Training]{\ref*{sec:Training}} explains the training process of EstNet and evaluates its estimation performance from multiple perspectives. In Sect. \hyperref[sec:Validation]{\ref*{sec:Validation}}, we compare and validate our work against the traditional parameter estimation methods used by Guo and Kepler \citep{guo2015white, kepler2021white}. Finally, the conclusions and future outlook are presented in Sect. \hyperref[sec:Conclusion]{\ref*{sec:Conclusion}}. 

\section{\textbf{Methods}}
\label{sec:Methods}
\subsection{EstNet Overview}
Our primary goal is to design an algorithm that can capture both the local spatial information of a specific spectral line and the long-range dependencies between different absorption lines. In astronomy, deep learning algorithms are mainly applied using CNN and RNN \citep{kattenborn2021review, sherstinsky2020fundamentals}. However, these algorithms cannot simultaneously extract both spatial and sequential information from spectra. CNN, through the action of convolutional kernels, only explores local regions of the input spectra, allowing the network to quickly focus on different characteristic absorption lines. However, different absorption lines have dependencies, and the key information in the spectra may be global and span multiple bands. To capture long-range dependencies in input data, sequence models represented by RNN can be used. However, RNN faces the core issue of long-term dependencies \citep{he2016deep}. During the learning process using backpropagation \citep{rojas1996backpropagation}, they may encounter gradient vanishing or explosion problems, making it difficult to model long-range dependencies in the input sequence.

Residual Network (ResNet), proposed by Kaiming He and colleagues from Microsoft Research Asia, has had a significant impact in the field of CNN \citep{he2016deep}. The residual connections in ResNet effectively address the network degradation problem, allowing the model to deepen and improve accuracy while ensuring convergence \citep{he2016identity}. This has garnered substantial attention from both academia and industry. ResNet has profoundly influenced subsequent network designs, and residual connections are now widely used in modern neural networks. In EstNet, each PRSE block in the PRSE module consists of two residual connections.

The memory capacity of neurons in RNN is limited. With the data being continuously fed into the model, previously accumulated information tends to be forgotten. An effective solution is to introduce gating mechanisms to control the accumulation of information in the hidden state, selectively adding new information and forgetting previously accumulated information, retaining only information valuable for prediction. Networks with such gating mechanisms include LSTM and GRU \citep{shewalkar2019performance}. In LSTM, the input gate, forget gate, and output gate work in unison to regulate the input, retention, and output of information, ensuring effective updating of the cell state. However, compared to the GRU, LSTM has a more complex structure with more parameters. In a GRU network, only the update gate and reset gate are used to balance the input of new information and the forgetting of irrelevant information. This simplifies the model structure and reduces computational complexity and memory requirements. This design improves training efficiency and often achieves performance better than LSTM in many applications. EstNet incorporates GRU cells to capture the global sequence information of spectra, thereby retaining memory that is beneficial for final predictions.

In designing the algorithm, we incorporated principles from the aforementioned algorithms and introduced attention mechanisms to enhance the focus on important spectral features. Inspired by the human nervous system, attention mechanisms selectively filter information and focus on what is important at a given moment \citep{niu2021review}. Currently, there are various types of attention mechanisms, including multi-head attention, channel attention, and spatial attention, which have shown excellent results in fields such as speech recognition, text generation, machine translation, and image classification \citep{he2016deep}. For spectral data, we need neural networks to focus more on important spectral features, such as hydrogen Balmer absorption lines, to enhance learning capabilities. Therefore, we introduced the SE attention mechanism after each residual connection in EstNet \citep{hu2018squeeze}.

EstNet integrates mainstream neural network design principles, combining various technologies into one neural network model. The backbone network incorporates CNN and RNN \citep{sherstinsky2020fundamentals, kattenborn2021review}. A fully connected feedforward neural network outputs the predicted labels \citep{sazli2006brief}. We proposed an adaptive loss mechanism and embedded Monte-Carlo Dropout to measure model uncertainty, providing confidence intervals for the estimates and enhancing model interpretability \citep{kendall2017uncertainties}. To increase the model's depth, we introduced residual connections in the PRSE blocks of EstNet, allowing the model to handle more complex estimation tasks \citep{he2016deep}. To prevent internal covariate shift, we included batch normalization layers after each convolutional layer \citep{bjorck2018understanding, ioffe2015batch}. Our training strategy employs weight decay and cosine annealing to better guide model convergence \citep{andriushchenko2023we, liu2022super}. To enhance the model's representation learning capability, we incorporated the SE attention mechanism in the PRSE module \citep{hu2018squeeze}. Next, we will introduce the main components of EstNet.

\begin{figure*}
    \centering
    \begin{subfigure}{0.48\textwidth}
        \centering
        \includegraphics[width=\textwidth]{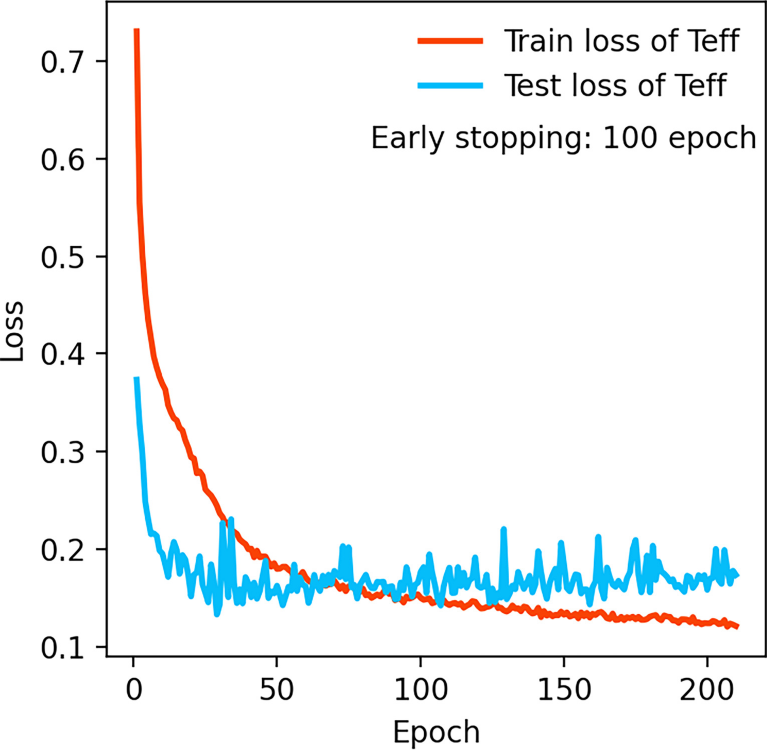}
        \caption{}
        \label{fig:teff_loss}
    \end{subfigure}%
    \begin{subfigure}{0.48\textwidth}
        \centering
        \includegraphics[width=\textwidth]{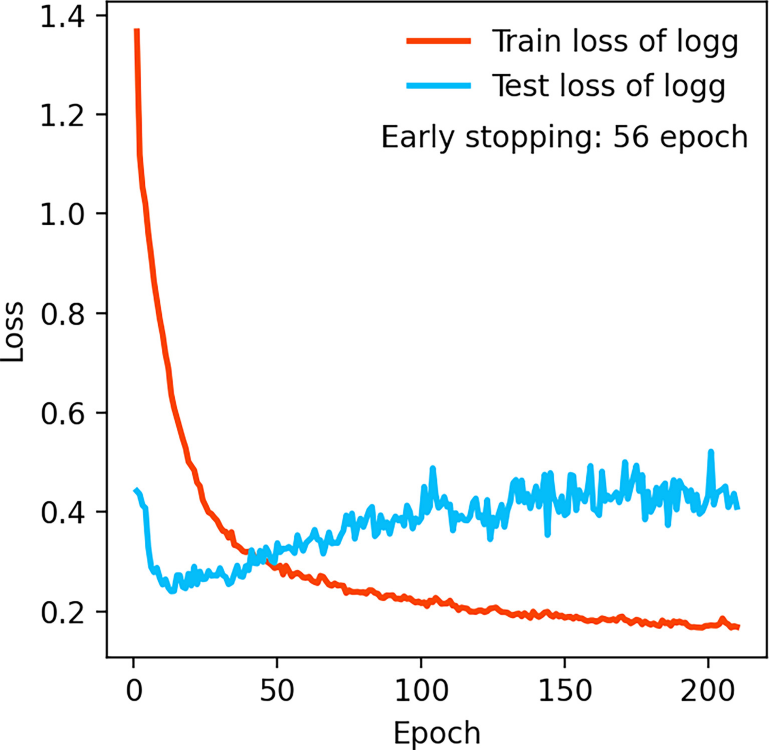}
        \caption{}
        \label{fig:logg_loss}
    \end{subfigure}
    \caption{Changes in loss during the training process. Left panel shows the loss changes during \(T_{\text{eff}}\) estimation, with early stopping at 100 epochs to obtain the optimal model parameters. Right panel shows the loss changes during \(\log\) g estimation, with early stopping at 56 epochs to obtain the optimal parameters. After each training epoch, EstNet loads the parameters from that epoch and validates them on a fixed test set, saving the parameters that achieve the best validation results across all epochs.}
    \label{fig:loss_changes}
\end{figure*}

\subsection{EstNet Structure}
EstNet consists of four modules: The Input Module, PRSE Module, GRU Module, and Output Module, as shown in Fig. \hyperref[fig:EstNet]{\ref*{fig:EstNet}}.

{\textbf{Input Module}}: ~~The input data \( x \) is a three-dimensional tensor with a shape of \((B, C, L)\), representing Batch size, Channel number, and Length, respectively. \( x \) passes through a convolutional layer with a kernel size of 3, which increases the input channels to 64 dimensions without changing the size of \( x \). Then it goes through a batch normalization (BN) layer and is activated by the PReLU function, resulting in the final output \( x_{B,64,L} \). The process is described in Eq. (\hyperref[eq1]{\ref*{eq1}}).
\begin{equation}
x_{B,64,L} = \text{input}(x_{B,1,L}) = \text{PReLU}(\text{BN}(\text{Conv}_3(x_{B,1,L})))
\label{eq1}
\end{equation}

{\textbf{PRSE Module}}: ~~The PRSE module is composed of multiple stacked PRSE blocks. Residual connections, first proposed in ResNet, address network degradation issues \citep{he2016deep}. The PRSE block incorporates two chained residual connections. The first residual connection outputs multi-channel feature maps, while the second applies different weights to each channel, enabling the network to focus on important channel information \citep{hu2018squeeze}. The process is detailed in Eq. (\hyperref[eq2]{\ref*{eq2}}).
\begin{equation}
x_{B,512,L/16} = \text{PRSE}(\text{PRSE}(\ldots \text{PRSE}(x_{B,64,L}) \ldots))
\label{eq2}
\end{equation}

{\textbf{GRU Module}}: ~~The GRU network is a type of RNN that introduces multiple gating mechanisms to effectively solve long-term dependency issues \citep{dey2017gate, bengio1993problem}. It models the dependencies of absorption lines at different wavelengths, extracting global sequence information from the entire spectrum. The input received by the GRU cell at each time step is downsampled by a factor of 16. By integrating information from different channels to capture global information, it ultimately produces a single-channel feature map \(x_{B,1,L/16}\), as shown in Eq. (\hyperref[eq3]{\ref*{eq3}}).
\begin{equation}
x_{B,1,L/16} = \text{GRU}(x_{B,512,L/16})
\label{eq3}
\end{equation}

{\textbf{Output Module}}: ~~The output module flattens the three-dimensional tensor \(x_{B,1,L/16}\) into a two-dimensional tensor \(x_{B,L/16}\), and then passes it through a fully connected layer to obtain the final output. A dropout layer is introduced before the fully connected layer, utilizing Monte-Carlo Dropout to deactivate hidden neurons with a certain probability, thereby enhancing the model's generalization ability \citep{kendall2017uncertainties}. This process is detailed in Eq. (\hyperref[eq4]{\ref*{eq4}}).
\begin{equation}
x_{B,1} = \text{Output}(x_{B,1,L/16}) = \text{FC}(\text{Flatten}(x_{B,1,L/16})) = \text{FC}(x_{B,L/16})
\label{eq4}
\end{equation}

\subsection{Adaptive Loss}

We designed an adaptive loss mechanism to specifically address low-resolution spectral data. EstNet, being a highly parameterized deep neural network model, possesses strong learning capabilities. However, during the first stage, samples that exceed the model's learning capacity often contain anomalous information, preventing the network from applying learned patterns effectively. Our adaptive loss mechanism enables EstNet to dynamically adjust its focus during the learning process. When dealing with challenging samples, the network automatically reduces attention to these anomalies and instead concentrates on normal samples. This strategy enhances the overall robustness and performance of the model. This approach to anomaly detection, based on neural networks \citep{han2006evolutionary}, is widely used in other fields as well. For example, LSTM and VAE have achieved significant success in anomaly detection in time series data \citep{ergen2019unsupervised, zhou2021vae, niu2020lstm}.

\begin{figure*}
    \centering
    \begin{subfigure}{0.48\textwidth}
        \centering
        \includegraphics[width=0.95\textwidth]{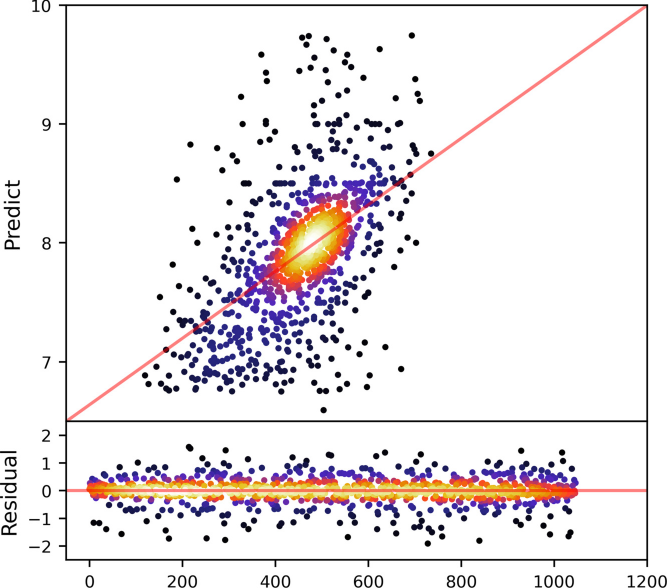}
        \caption{}
        \label{fig:logg}
    \end{subfigure}%
    \begin{subfigure}{0.48\textwidth}
        \centering
        \includegraphics[width=\textwidth]{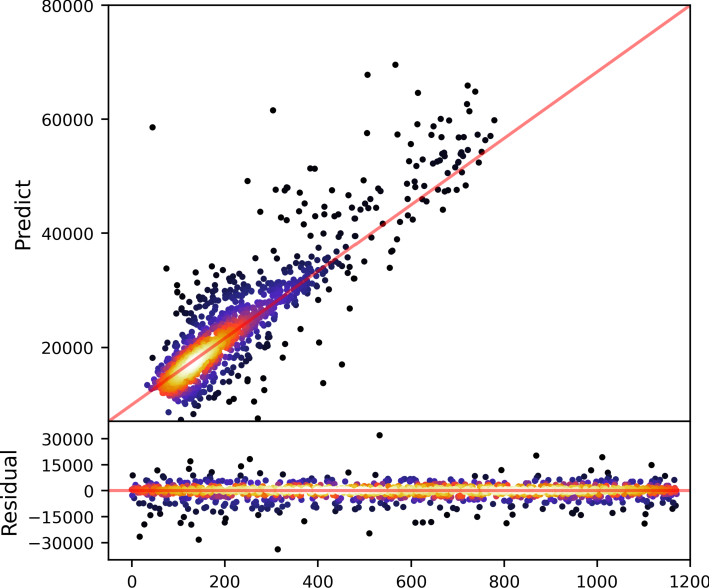}
        \caption{}
        \label{fig:Teff}
    \end{subfigure}
    \caption{(a) Estimation results for \(\log\) g. (b) Estimation results for \(T_{\text{eff}}\). The upper subplots in both figures show the kernel density plots of predicted and true labels, with yellow areas indicating high-density regions. The lower subplots show the residual distribution between predicted and true labels. The horizontal axis represents sample point indices, and the vertical axis represents residuals and predicted labels. The red solid line represents the identity line.}
    \label{fig:predicton}
\end{figure*}

In the first stage, EstNet measures the anomaly degree of each sample. In the second stage, EstNet loads the anomaly information obtained from the first stage, then performs limited learning on the anomalous samples and extensive learning on the non-anomalous samples.

$D = \left\{(x^{(n)}, y^{(n)})\right\}_{n=1}^{N}$ is a dataset of size \(N\). $D_{\text{train}} = \left\{(x^{(n)}, y^{(n)})\right\}_{n=1}^{O}$ is the training set, where \(O\) is the training set size. $f_{\theta}(x) = f(x, \theta)$ refers to the EstNet network used in our study, which can be considered a nonlinear function. $f_{\theta}(x) \in \mathcal{F}$, where \(\mathcal{F}\) is a set of functions. \(\theta \in \Omega\), \(\theta\) is a learnable parameter, and \(\Omega\) is the parameter space.

In the first phase, we consider the loss function \( L^{\text{step1}}(y, f_{\theta}(x)) \) defined in Eq. (\hyperref[eq5]{\ref*{eq5}}), where \( \delta \) is a small hyperparameter. When the residual is less than or equal to \( \delta \), \( L^{\text{step1}} \) reduces to \( L_2 \) loss. Otherwise, \( L^{\text{step1}} \) reduces to \( L_1 \) loss, ensuring the model's differentiability at the origin.

\begin{equation}
L^{\text{step1}}(y, f_{\theta}(x)) = 
\begin{cases} 
\frac{1}{2} (y - f_{\theta}(x))^2 & \lvert y - f_{\theta}(x) \rvert \leq \delta \\
\delta \lvert y - f_{\theta}(x) \rvert - \frac{1}{2} \delta^2 & \lvert y - f_{\theta}(x) \rvert > \delta 
\end{cases}
\label{eq5}
\end{equation}

The parameters in \( f_{\theta}(x) \) are initialized using a normal distribution, \(\theta \sim N(0, \sigma^2)\). The learning algorithm employs Mini-Batch Gradient Descent \citep{bottou2010large}, which randomly selects a small subset of training samples during each iteration to compute the gradients and update the parameters. In the \( t \)-th iteration, the algorithm randomly selects a subset \(\gamma_t\) containing \( K \) samples and computes the average gradient of the loss function for each sample in this subset. The parameters are updated using Eq. (\hyperref[eq6]{\ref*{eq6}}), with \(\alpha\) as the learning rate.

\begin{equation}
\theta_{t+1} \leftarrow \theta_{t} - \alpha \frac{1}{K} \sum_{(x,y) \in \gamma_{t}} \frac{\partial L^{\text{step1}}(y, f(x; \theta))}{\partial \theta}
\label{eq6}
\end{equation}

Given the threshold \(\varepsilon\), the model stops learning when Eq. (\hyperref[eq7]{\ref*{eq7}}) is satisfied, and the learned parameters \(\theta^{\text{step1}}\) are saved.

\begin{equation}
|L(y, f_{\theta}(x))| < \varepsilon, \forall x \in D_{\text{train}}
\label{eq7}
\end{equation}

By loading the parameters \(\theta^{\text{step1}}\) from the first stage, we obtain the model \(f_{\theta^{\text{step1}}}(x)\). Using Eq. (\hyperref[eq8]{\ref*{eq8}}), we calculate the absolute distance \(\alpha(y \mid x)\) between the predicted and true values for each sample in \(D_{\text{train}}\). Equation (\hyperref[eq9]{\ref*{eq9}}) normalizes the absolute distance to obtain the anomaly score \(\beta(y \mid x)\) for each sample.

\begin{equation}
\alpha(y \mid x) = \lvert y - f_{\theta^{\text{step1}}}(x) \rvert
\label{eq8}
\end{equation}

\begin{equation}
\beta(y \mid x) = \frac{\alpha(y \mid x) - \alpha_{\min}}{\alpha_{\max} - \alpha_{\min}}
\label{eq9}
\end{equation}

In the second stage, we employ Eq. (\hyperref[eq10]{\ref*{eq10}}) to perform kernel density estimation on the labels \( \{ y^{(n)} \}_{n=1}^{O} \), where \(O\) is the size of the training set. Our model employs the Gaussian kernel function \(K(y)\), as shown in Eq. (\hyperref[eq11]{\ref*{eq11}}).

\begin{equation}
G_{0}(y) = \frac{1}{O} \sum_{i=1}^{O} \frac{1}{H} K \left( \frac{y^{i} - y}{H} \right)
\label{eq10}
\end{equation}

\begin{equation}
K(y) = \frac{1}{\sqrt{2\pi}} \exp\left(-\frac{1}{2} y^{2}\right)
\label{eq11}
\end{equation}

Using the parameters \(\theta^{\text{step1}}\) from the first stage as the initial parameters for the second stage reduces the training time. The second stage loss function is adaptively adjusted based on \(\beta(y \mid x)\) obtained from the first stage, ensuring minimal learning on anomalous data and extensive learning on non-anomalous data. The factor \(\mu\) balances the relative importance of \(G(y)\) and \(\beta(y \mid x)\). The loss function for the second stage, \(L^{\text{step2}}\), is defined as:

\begin{figure*}
    \centering
    \begin{subfigure}{0.48\textwidth}
        \centering
        \includegraphics[width=0.7\textwidth]{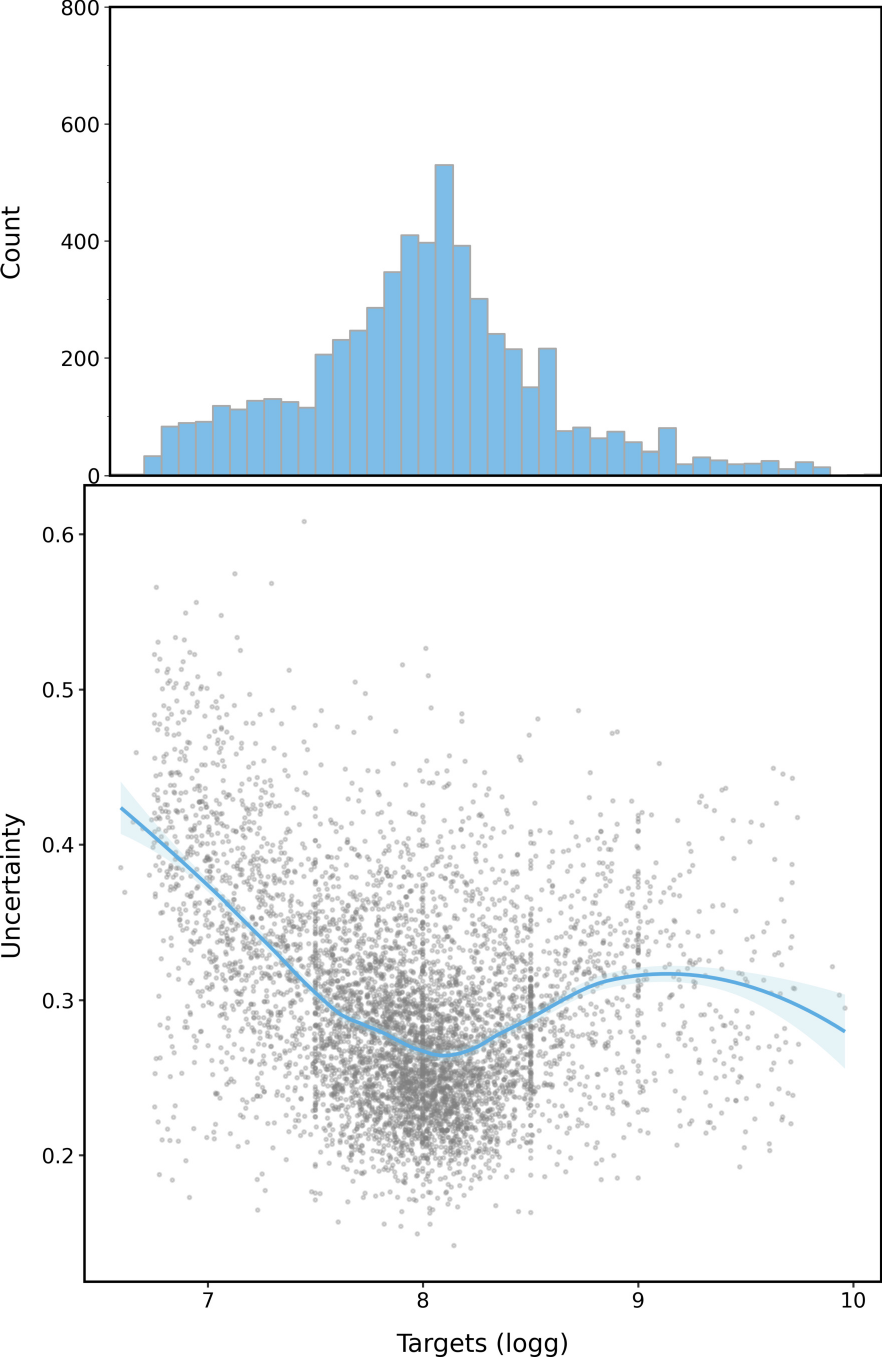}
        \caption{}
        \label{fig:mc_logg}
    \end{subfigure}%
    \begin{subfigure}{0.48\textwidth}
        \centering
        \includegraphics[width=0.7\textwidth]{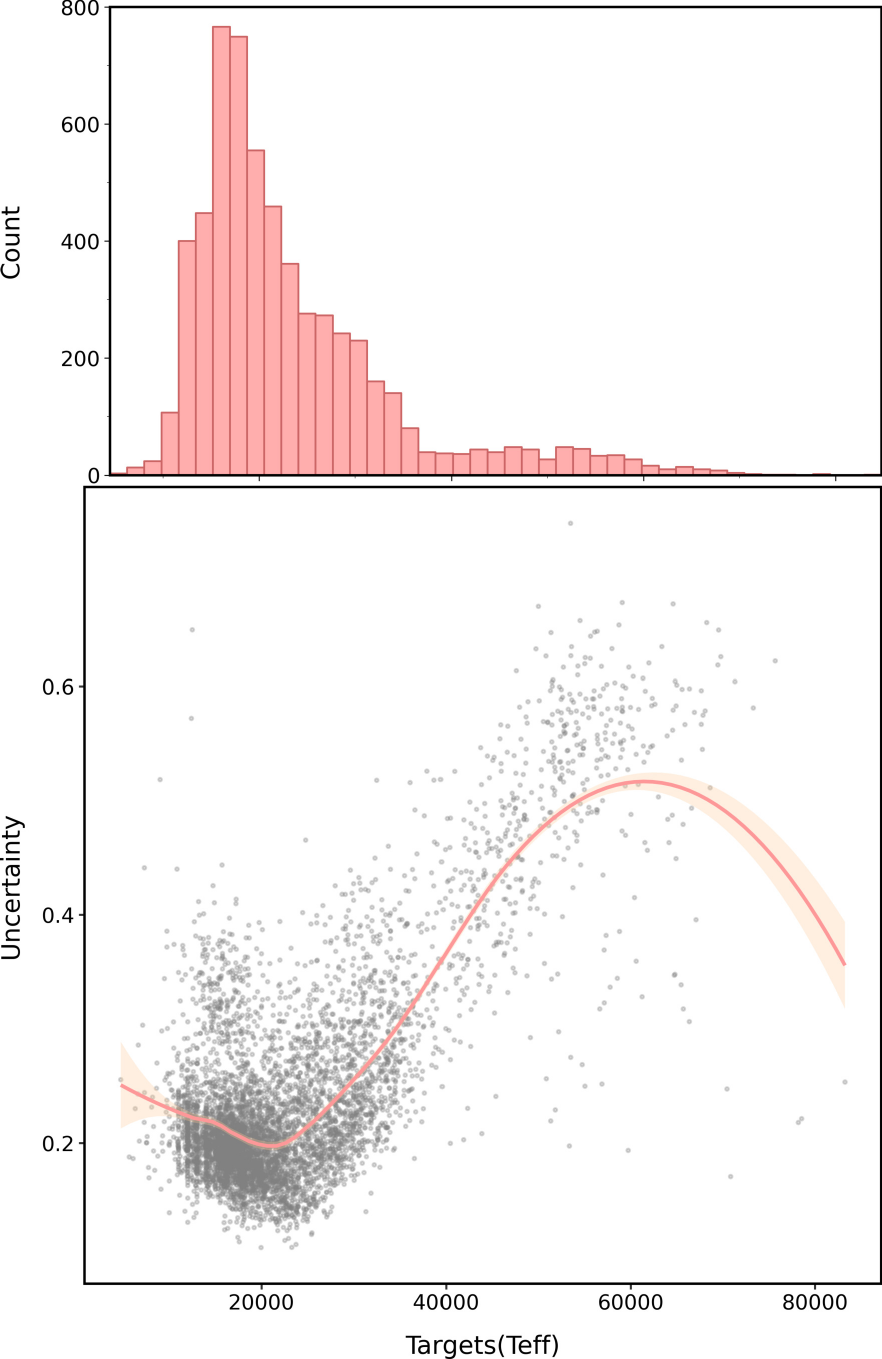}
        \caption{}
        \label{fig:mc_Teff}
    \end{subfigure}
    \caption{Reliability analysis for \(\log\) g (left panel) and \(T_{\text{eff}}\) (right panel). The smooth curves represent uncertainty trend of the labels, and the shaded areas correspond to 95\% confidence intervals.}
    \label{fig:mc_dropout}
\end{figure*}

\begin{equation}
L^{\text{step2}}(y, f_{\theta^{\text{step1}}}(x)) = 
\begin{cases}
\frac{1}{2} G(y) 10^{\mu}(1 - \beta(y \mid x))(y - f_{\theta^{\text{step1}}}(x))^2 \\
\delta G(y) 10^{\mu}(1 - \beta(y \mid x))|y - f_{\theta^{\text{step1}}}(x)| - \frac{1}{2}\delta^2 
\end{cases}
\end{equation}

\noindent
Mini-batch gradient descent is then used again as the learning algorithm to optimize \(L^{\text{step2}}\) and obtain the optimal parameters \(\theta^{\text{opt}}\), resulting in the optimal model \(f_{\theta^{\text{opt}}}(x)\):

\begin{equation}
f_{\theta^{\text{opt}}}(x) = \arg\min_{\theta \in \mathcal{F}} L^{\text{step2}}(y, f_{\theta^{\text{step1}}}(x))
\end{equation}

\subsection{Monte-Carlo Dropout Layer}

Monte Carlo Dropout is commonly employed to estimate the uncertainty of model predictions in neural networks. This technique involves performing multiple forward passes during inference, with different neurons randomly dropped each time. The results of these multiple passes are then statistically analyzed to estimate the uncertainty in the output \citep{gal2016dropout}. This uncertainty encompasses both the inherent data noise and the uncertainty arising from the parameters of the model \citep{kendall2017uncertainties}. EstNet integrates Monte Carlo Dropout in its output module to evaluate prediction confidence, thereby enhancing the understanding of its predictive behavior across different regions.

Let \(Z_1 = \{Z_{i,j}\}_{Q \times Q} \in \mathbb{R}^{Q \times Q}\) and \(Z_2 = \{Z_{i,j}\}_{K \times K} \in \mathbb{R}^{K \times K}\) be random matrices. Each element of \(Z_1\) and \(Z_2\) follows a Bernoulli distribution, \(Z_{i,j} \sim \text{Bernoulli}(P_{i,j})\). The final network output can be represented as \(\hat{y} = \text{Sigmoid}(X(Z_1W_1) + m)(Z_2W_2) \). Here, \( X \in R^{1 \times Q} \) is the input feature vector of the Output module, \( W_1 \in R^{Q \times K} \) and \( W_2 \in R^{K \times K} \) are learnable parameters, and \( m \in R^{1 \times K} \) is the bias vector.

For \(T\) forward passes, the final estimate is given by Eq. (\hyperref[eq14]{\ref*{eq14}}), and the uncertainty measurement \( \text{Var}(\hat{y}_1, \ldots, \hat{y}_T) \) is given by Eq. (\hyperref[eq15]{\ref*{eq15}}). Here, \( W_1^t \) and \( W_2^t \) are the parameter matrices for the \( t \)-th iteration, and \( \hat{y}_t(x \mid W_1^t, W_2^t) \) denotes the outcome of the \( t \)-th forward pass, with \( \hat{y}_t \in R^{1 \times K} \).

\begin{equation}
\hat{y} = E(y \mid x) \approx T^{-1} \sum_{t=1}^{T} \hat{y}_t(x \mid W_1^t, W_2^t)
\label{eq14}
\end{equation}
\begin{equation}
\text{Var}(\hat{y}_1, \ldots, \hat{y}_T) \approx (T - 1)^{-1} \sum_{t=1}^{T} (\hat{y}_i - \hat{y})^T (\hat{y}_i - \hat{y})
\label{eq15}
\end{equation}

The Monte-Carlo Dropout layer provides accurate parameter estimates for white dwarfs and measures the uncertainty of these estimates, allowing for interval estimation of the results. These procedures enhance the robustness and interpretability of the model.

\subsection{Evaluation Metrics}

To comprehensively evaluate the performance of EstNet, we selected MAE, RMSE, MAPE, and ME as metrics for assessing model error. Smaller values of these metrics indicate lower prediction errors and higher accuracy in the model's predictions. The calculation methods for these four evaluation metrics are as follows:

\begin{equation}
\text{MAE} = M^{-1} \sum_{i=1}^{M} \left| f_{\theta_{\text{opt}}}(x_i) - y_i \right|
\end{equation}
\begin{equation}
\text{RMSE} = \sqrt{M^{-1} \sum_{i=1}^{M} \left( f_{\theta_{\text{opt}}}(x_i) - y_i \right)^2}
\end{equation}
\begin{equation}
\text{MAPE} = M^{-1} \sum_{i=1}^{M} \left| \frac{f_{\theta_{\text{opt}}}(x_i) - y_i}{y_i} \right|
\end{equation}
\begin{equation}
\text{ME} = \text{Median} \left( \left| f_{\theta_{\text{opt}}}(x_i) - y_i \right| \right)
\end{equation}

Here, \(f_{\theta_{\text{opt}}}(x)\) is the optimal model obtained after training. $D_{\text{test}} = \left\{ (x^{(n)}, y^{(n)}) \right\}_{n=1}^{M}$ represents the test set of size \(M\) selected from \(\mathcal{D}\). $x_i$ denotes the features of the \(i\)-th sample, and $y_i$ represents the true label of the \(i\)-th sample.

\section{\textbf{Data}}
\label{sec:Data}
The LAMOST telescope, situated at the Xinglong Observatory of the National Astronomical Observatories, Chinese Academy of Sciences, is renowned for its significant aperture and extensive field of view \citep{su1998large}. It incorporates a unique Schmidt reflective design and is equipped with 4000 optical fibers on its focal plane, enabling the simultaneous observation of 4000 targets over a wide area. This configuration significantly enhances the efficiency of spectral acquisition \citep{yao2012site}. The LAMOST survey, in its eighth data release (DR8), has provided spectra for over 10 million stars, approximately 220,000 galaxies, and about 71,000 quasars \citep{yan2022overview}. The extensive spectral data provided by LAMOST has significantly advanced our understanding and exploration of the Milky Way and its millions of stars.

\begin{figure*}
    \centering
    \begin{subfigure}{0.33\textwidth}
        \centering
        \includegraphics[width=\textwidth]{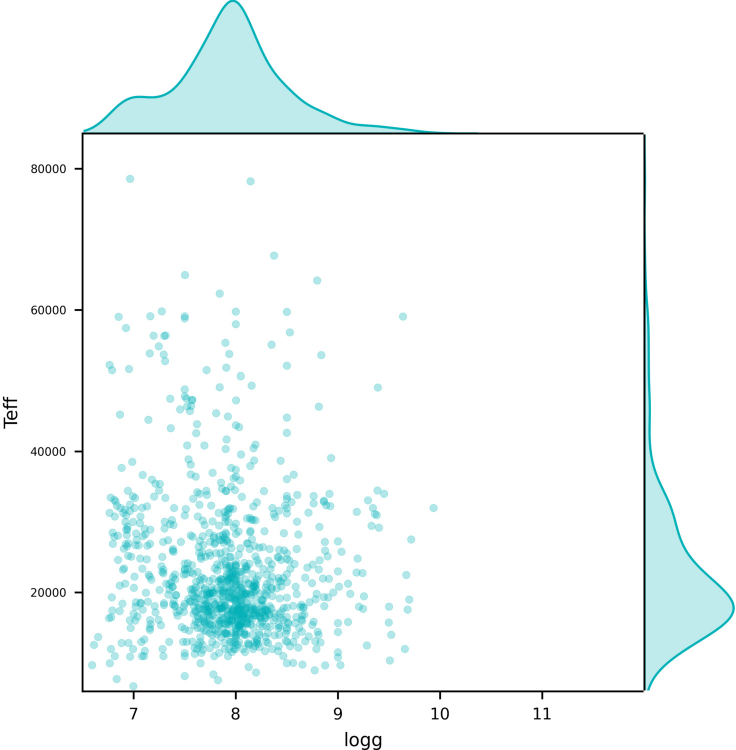}
        \caption{}
        \label{fig:before_noise}
    \end{subfigure}%
    \begin{subfigure}{0.33\textwidth}
        \centering
        \includegraphics[width=\textwidth]{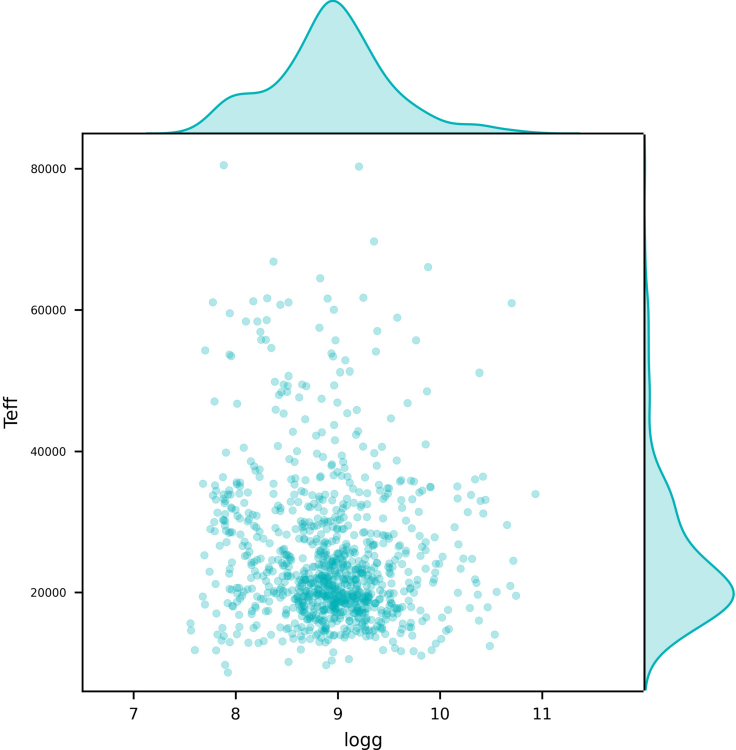}
        \caption{}
        \label{fig:after_noise}
    \end{subfigure}%
    \begin{subfigure}{0.33\textwidth}
        \centering
        \vspace{0.55cm}
        \includegraphics[width=\textwidth]{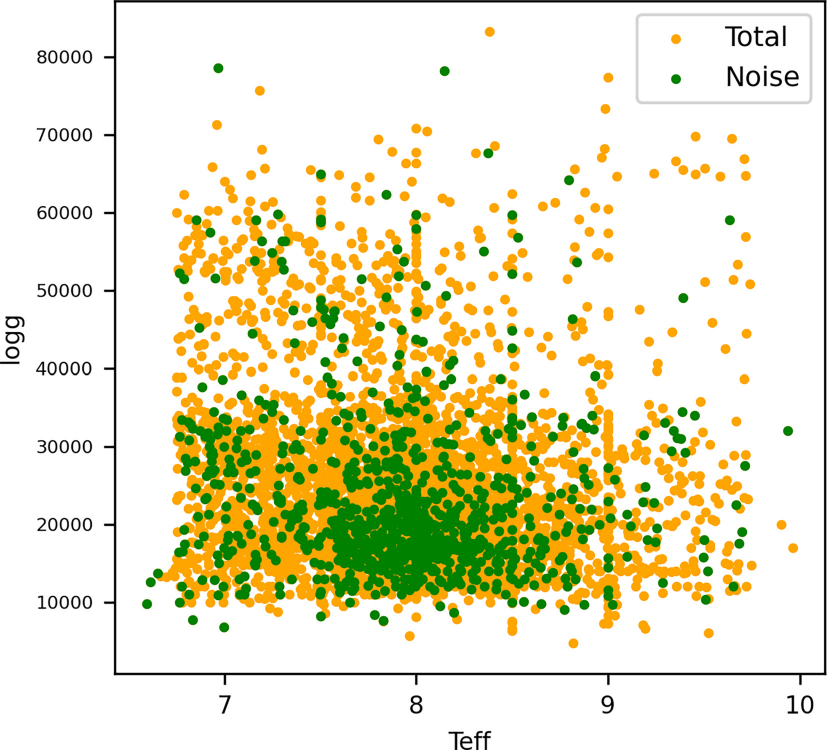}
        \caption{}
        \label{fig:SN}
    \end{subfigure}
    \caption{Noise Processing Figure. (a) shows the label distribution before adding noise. (b) shows the label distribution after adding noise. (c) depicts the distribution of the 954 data points with added noise (green circles) within the original dataset of 5,965 data points (yellow circles).}
    \label{fig:noise_processing}
\end{figure*}

\citet{gentile2021catalogue} utilized Gaia EDR3 to identify 1.3 million white dwarf candidates based on their absolute magnitude, color, and mass. \citet{kong2021identification} cross-matched these Gaia EDR3 white dwarf candidates with LAMOST DR7 data within a radius of 3 arcseconds, obtaining 12,046 corresponding LAMOST spectra. Subsequently, they employed a support vector machine algorithm to further identify 9,496 white dwarf candidates. After manual spectral verification, they confirmed 6,190 white dwarfs. Using a template matching method, they estimated the physical atmospheric parameters (\(\log\) g, \(T_{\text{eff}}\)) of the white dwarfs, ultimately providing a catalog of stars with parameters.

Using the catalog provided by Kong, we downloaded low-resolution (R$\sim$1800) spectral data from LAMOST. Each spectrum covers the wavelength range of 4000 \AA{}-8000 \AA{} and contains 3909 feature points. After removing data without parameters or with missing signal-to-noise ratios, we obtained 5965 spectra for model training and testing. 

The characteristic absorption lines of white dwarfs are primarily concentrated in the blue end \citep{kong2018spectral}. In our experiment, the data quality was extremely poor, as shown in Fig. \hyperref[fig:predicton]{\ref*{fig:predicton}}. The low-resolution, low signal-to-noise ratio spectral data exhibits issues such as flux loss and excessive noise. Compared to most studies that use high signal-to-noise ratio data (SN > 10) for parameter estimation, our task of conducting parameter estimation based on poor-quality data is highly challenging. During actual training, we do not perform any interpolation for the missing values of the features. This is because our model can adaptively learn from incomplete data, highlighting one of the key advantages of our design.

\section{\textbf{EstNet Training and Evaluation}}
\label{sec:Training}
\subsection{Estimation of \(T_{\text{eff}}\) and \(\log\) g}
During parameter estimation, we divided 5,965 spectral data samples into training and test sets in an 8:2 ratio. Depending on the number of stacked PRSE blocks, we can define EstNet models of different depths. For white dwarfs, the main indicator of \(\log\) g is the width of the atmospheric absorption lines \citep{kepler2021white}, while \(T_{\text{eff}}\) is related to the flux intensity at different wavelengths \citep{prokhorov2009calculation}. EstNet finds it more challenging to capture the width information of absorption lines than the flux intensity information. Therefore, we use the EstNet66 model, which has a larger number of parameters, to estimate \(\log\) g, and the EstNet34 model to estimate \(T_{\text{eff}}\).

For \(T_{\text{eff}}\) estimation, we use EstNet34, which includes 34 convolutional layers. During training, a 0.6 dropout rate is applied to the fully connected layers, with a weight decay of 0.0005. For \(\log\) g estimation, we use EstNet66, which includes 66 convolutional layers. During training, a 0.7 dropout rate is applied to the fully connected layers, with a weight decay of 0.0015. The changes in loss during the training process are shown in Fig. \hyperref[fig:loss_changes]{\ref*{fig:loss_changes}}.

As shown in Fig. \hyperref[fig:predicton]{\ref*{fig:predicton}}, we plot the distribution of predicted and true labels. The scatter points are concentrated around the identity line, indicating that the predicted values are close to the true values. Specifically, the MAE, RMSE, MAPE, and ME for \(\log\) g are 0.31 dex, 0.46 dex, 3.97\%, and 0.19 dex, respectively. For \(T_{\text{eff}}\), the MAE, RMSE, MAPE, and ME are 3128 K, 5062 K, 14.86\%, and 1826 K, respectively.

\begin{figure*}
    \centering
    \begin{subfigure}{0.48\textwidth}
        \centering
        \includegraphics[width=0.8\textwidth]{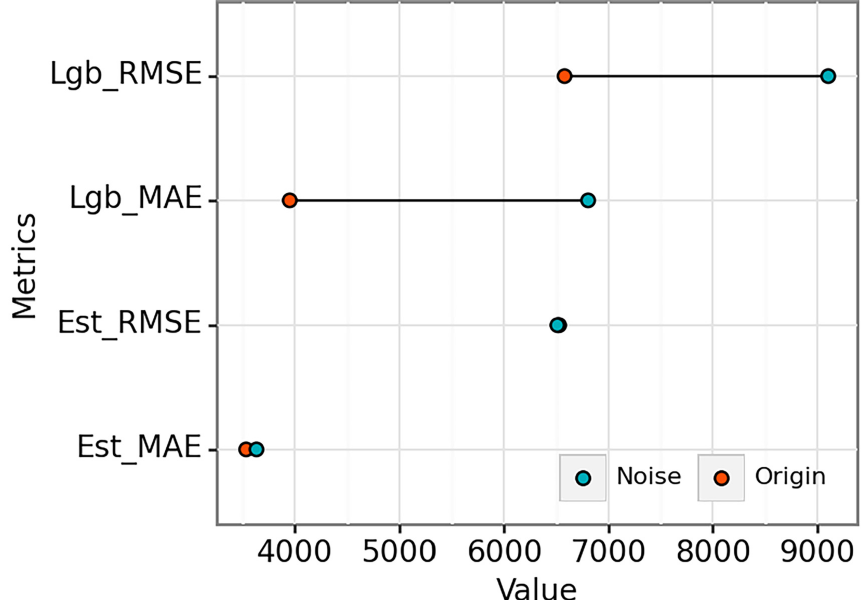}
        \caption{}
        \label{fig:subfig1_lgb}
    \end{subfigure}
    \begin{subfigure}{0.48\textwidth}
        \centering
        \includegraphics[width=0.8\textwidth]{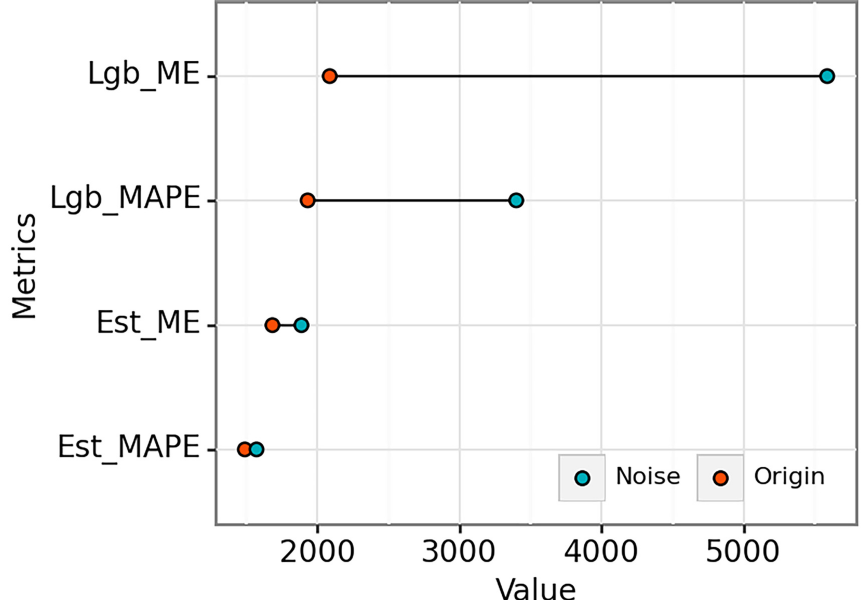}
        \caption{}
        \label{fig:subfig2_lgb}
    \end{subfigure}

    \vspace{1em} 

    \begin{subfigure}{0.48\textwidth}
        \centering
        \includegraphics[width=0.8\textwidth]{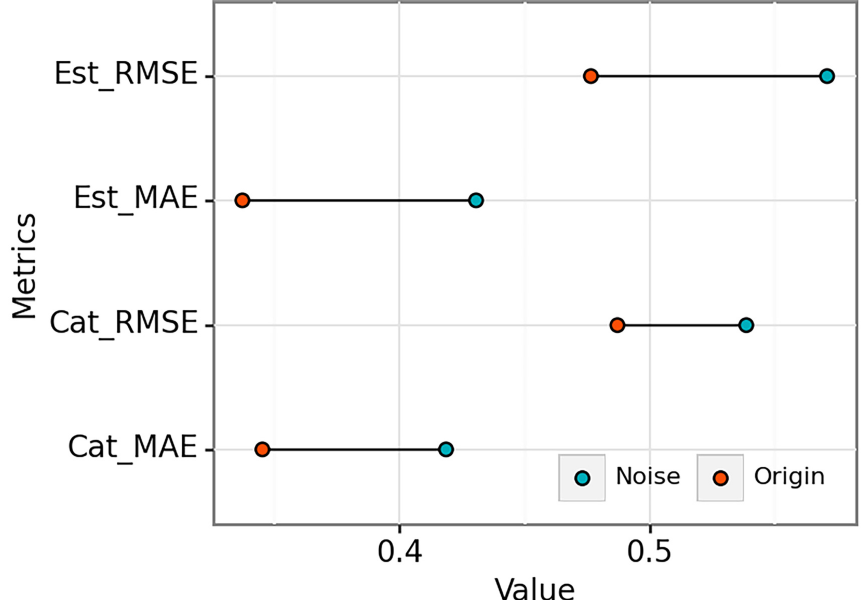}
        \caption{}
        \label{fig:subfig3_cat}
    \end{subfigure}
    \begin{subfigure}{0.48\textwidth}
        \centering
        \includegraphics[width=0.8\textwidth]{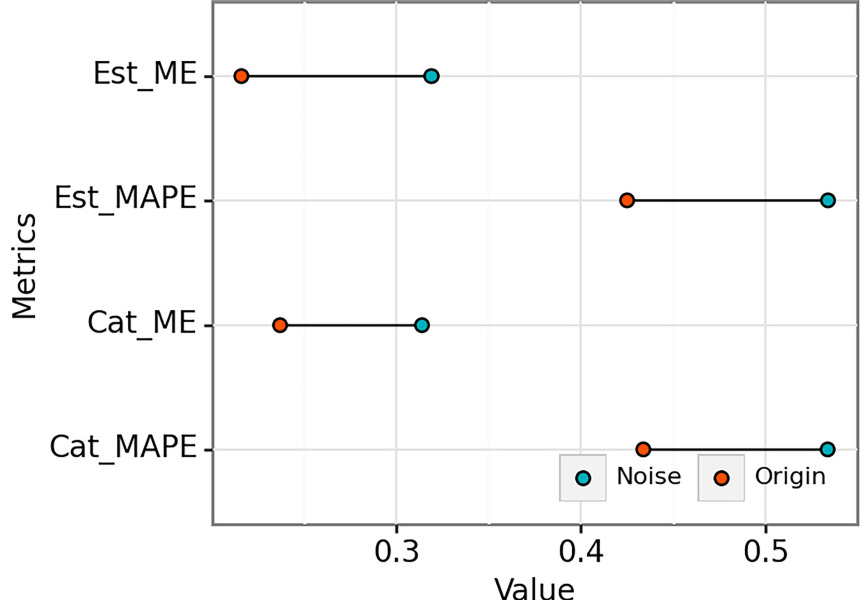}
        \caption{}
        \label{fig:subfig4_cat}
    \end{subfigure}
    \caption{Dumbbell plots of metric fluctuations before and after adding noise. (a) shows the fluctuations in RMSE and MAE for \(T_{\text{eff}}\) between LightGBM and EstNet34. (b) shows the fluctuations in MAPE and ME for \(T_{\text{eff}}\) between LightGBM and EstNet34. (c) shows the fluctuations in RMSE and MAE for \(\log\) g between CatBoost and EstNet66. (d) shows the fluctuations in MAPE and ME for \(\log\) g between CatBoost and EstNet66. The red points represent the metrics before adding noise, and the blue points represent the metrics after adding noise. The lengths of the lines indicate the degree of fluctuation in the evaluation metrics before and after adding noise. In (b) and (d), the MAPE values have been scaled to match the magnitude of ME for consistency in the plots.}
    \label{fig:combined_lgb_cat}
\end{figure*}

\subsection{Reliability Analysis}

The Monte Carlo Dropout layer in EstNet allows us to measure uncertainty in predictions for  \(T_{\text{eff}}\) and \(\log\) g, thereby helping us analyze the reliability of the model's outputs. As shown in Fig. \hyperref[fig:mc_dropout]{\ref*{fig:mc_dropout}}, the solid line represents the uncertainty trend, and the shaded area corresponds to the 95\% confidence interval. When predictions approach the boundary of training data, the prediction fluctuations of EstNet become more pronounced due to the smaller sample size, resulting in wider confidence intervals and greater sensitivity to the data.

The effective temperature of white dwarfs is concentrated between 4000 K and 150,000 K, with an average surface gravity acceleration of \(10^8 \, \text{cm} \, \text{s}^{-2}\) (\(\log\) g = 8) \citep{fontaine2001potential}. When \(\log\) g is close to 8 and \(T_{\text{eff}}\) ranges from 10,000 K to 50,000 K, the confidence interval narrows, indicating lower uncertainty in the predictions. This suggests that EstNet's predictions align well with the actual parameters of white dwarfs. When predictions closely match the true labels, EstNet provides narrower confidence intervals, supporting researchers in making reliable decisions based on the model's output.

\subsection{Comparative Analysis}
\label{sec:comparative}
With the advancement of machine learning, numerous algorithms have been applied to various astronomical tasks. However, there is still relatively little research on using machine learning algorithms to estimate white dwarf parameters. We compare the EstNet with other algorithms, including Random Forest \citep{breiman2001random}, XGBoost \citep{chen2016xgboost}, NGBoost \citep{duan2020ngboost}, CatBoost \citep{prokhorenkova2018catboost}, and LightGBM \citep{ke2017lightgbm}.

   \begin{table}
      \caption{Comparison with other models on \(T_{\text{eff}}\)}
         \label{model_comparison_teff}
     $$
         \begin{array}{lcccc}
            \hline\hline
            \noalign{\smallskip}
            \text{Model} & \text{MAE(K)} & \text{RMSE(K)} & \text{MAPE(\%)} & \text{M(K)} \\
            \noalign{\smallskip}
            \hline
            \noalign{\smallskip}
            \text{RF} & 4217 & 6686 & 19.80\% & 2480 \\
            \text{NGBoost} & 3992 & 6541 & 18.48\% & 2361 \\
            \text{CatBoost} & 4264 & 6629 & 19.74\% & 2831 \\
            \text{XGBoost} & 4038 & 7015 & 18.46\% & 2175 \\
            \text{LightGBM} & 3950 & 6895 & 18.30\% & 2082 \\
            \text{EstNet34} & 3128 & 5062 & 14.86\% & 1826 \\
            \noalign{\smallskip}
            \hline
         \end{array}
     $$
   \end{table}

   \begin{table}
      \caption{Comparison with other models on \(\log\) g}
         \label{model_comparison_logg}
     $$
         \begin{array}{lcccc}
            \hline\hline
            \noalign{\smallskip}
            \text{Model} & \text{MAE(dex)} & \text{RMSE(dex)} & \text{MAPE(\%)} & \text{M(dex)} \\
            \noalign{\smallskip}
            \hline
            \noalign{\smallskip}
            \text{RF} & 0.37 & 0.50 & 4.64\% & 0.26 \\
            \text{NGBoost} & 0.36 & 0.50 & 4.55\% & 0.24 \\
            \text{CatBoost} & 0.34 & 0.48 & 4.31\% & 0.21 \\
            \text{XGBoost} & 0.36 & 0.50 & 4.55\% & 0.24 \\
            \text{LightGBM} & 0.40 & 0.54 & 5.00\% & 0.28 \\
            \text{EstNet66} & 0.31 & 0.46 & 3.97\% & 0.19 \\
            \noalign{\smallskip}
            \hline
         \end{array}
     $$
   \end{table}

Random Forest has been widely used in the field of astronomy \citep{torres2019random, echeverry2022random, chandra2020computational, guo2022white}. The new members of the Boosting family, LightGBM and CatBoost, also exhibit excellent performance. However, our EstNet shows significantly stronger learning capabilities than these models, excelling across all evaluation metrics. As shown in Table \hyperref[model_comparison_teff]{\ref*{model_comparison_teff}}, for \(T_{\text{eff}}\) estimation, the two best models are EstNet66 and LightGBM, with EstNet66 achieving a mean absolute percentage error (MAPE) 3.44\% lower than LightGBM. As shown in Table \hyperref[model_comparison_logg]{\ref*{model_comparison_logg}}, for \(\log\) g estimation, the two best models are EstNet34 and CatBoost, with EstNet34 achieving a 0.34\% lower MAPE than CatBoost.

\begin{figure*}
    \centering
    \begin{subfigure}{0.48\textwidth}
        \centering
        \includegraphics[width=0.8\textwidth]{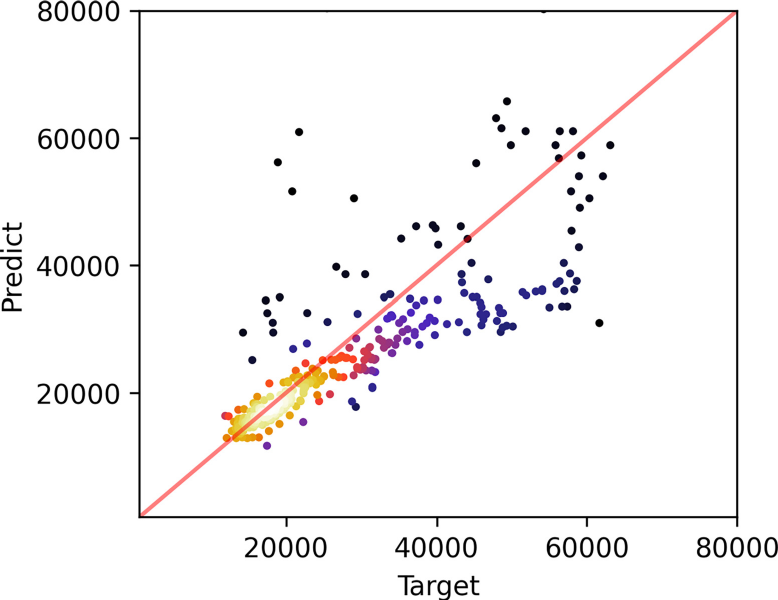}
        \caption{}
        \label{fig:teff_guo}
    \end{subfigure}
    \begin{subfigure}{0.48\textwidth}
        \centering
        \includegraphics[width=0.8\textwidth]{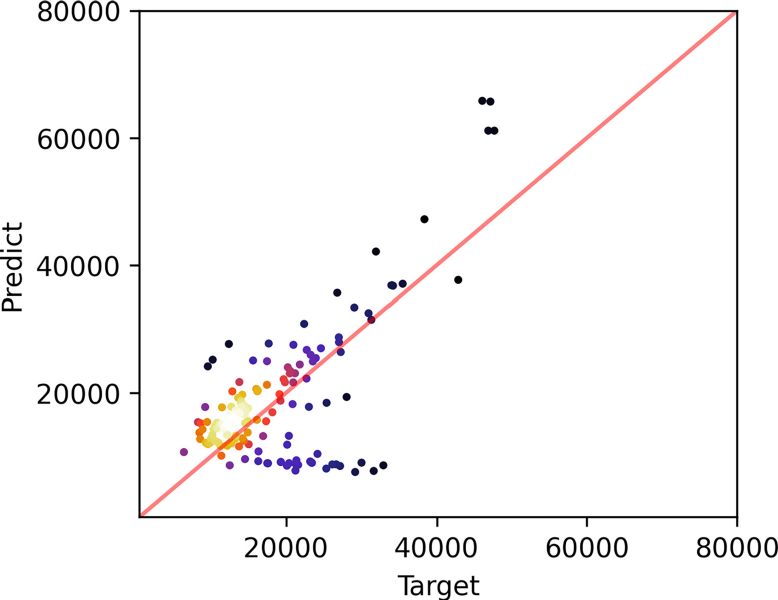}
        \caption{}
        \label{fig:teff_kepler}
    \end{subfigure}

    \vspace{1em} 

    \begin{subfigure}{0.48\textwidth}
        \centering
        \includegraphics[width=0.7\textwidth]{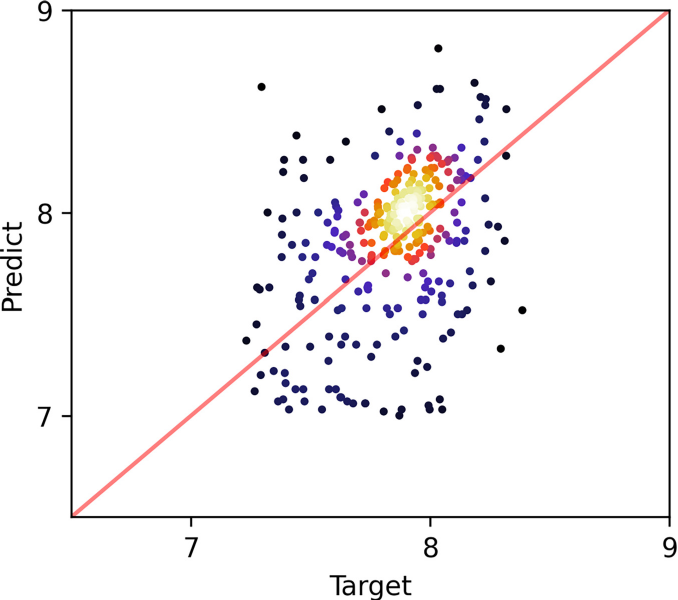}
        \caption{}
        \label{fig:logg_guo}
    \end{subfigure}
    \begin{subfigure}{0.48\textwidth}
        \centering
        \includegraphics[width=0.7\textwidth]{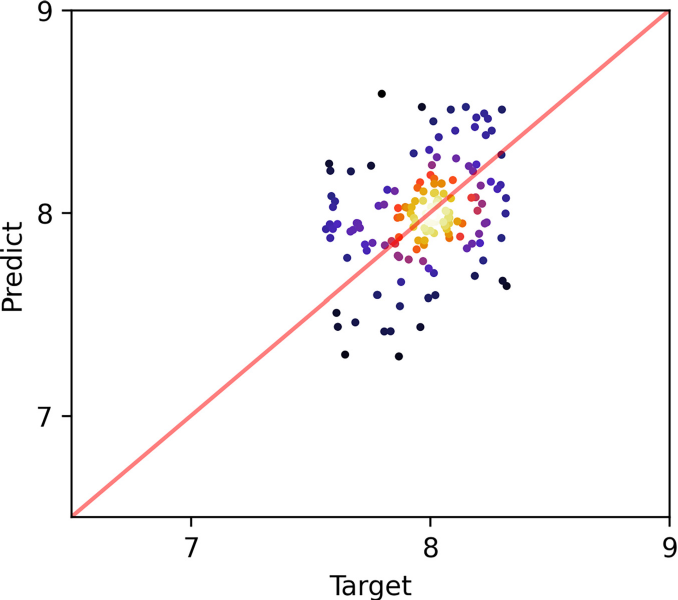}
        \caption{}
        \label{fig:logg_kepler}
    \end{subfigure}
    \caption{Comparison with traditional methods in predicting \(\log\) g and \(T_{\text{eff}}\) of white dwarfs. (a) and (b) show the predictions for \(T_{\text{eff}}\), while (c) and (d) display the predictions for \(\log\) g. (a) and (c) compare our results with those of Guo et al., using data from LAMOST. (b) and (d) compare our results with those of Kepler et al., using data from SDSS. The scatter point density in the yellow region is higher than in the purple region. The red solid line represents the identity line. The scatter points are distributed around the identity line, indicating that the predictions of EstNet are consistent and effective compared to traditional methods.}
    \label{fig:kepler_guo}
\end{figure*}

\subsection{Robustness Analysis}

Our model is data-driven, and our data has both low resolution and a low signal-to-noise ratio. The spectral data is very noisy, with issues such as missing flux and anomalous labels. However, with its inherent adaptive loss mechanism, EstNet can automatically reduce learning from anomalous data and increase learning from normal data.

To demonstrate EstNet's robustness to noisy data, we further compare it with the suboptimal models from Sect. \hyperref[sec:comparative]{\ref*{sec:comparative}}, LightGBM, and CatBoost. We divided 5,965 spectral data points into a training set containing 4,772 spectra using an 8:2 ratio. Within the training set, we applied noise to the labels of 954 spectra, also using an 8:2 ratio. For \(T_{\text{eff}}\) labels, the noise follows a normal distribution with a mean of 2000 and a standard deviation of 100. For \(\log\) g labels, the noise follows a normal distribution with a mean of 1 and a standard deviation of 0.1. Comparing parts (a) and (b) of Fig. \hyperref[fig:noise_processing]{\ref*{fig:noise_processing}}, the distribution of labels changes significantly after adding noise. For \(\log\) g, the original data was concentrated around 8 dex, and after adding noise, it shifts to around 9 dex.

Figure \hyperref[fig:combined_lgb_cat]{\ref*{fig:combined_lgb_cat}} shows a dumbbell plot illustrating the fluctuations in various metrics before and after adding noise. Before adding noise, the errors of EstNet are smaller than those of LightGBM and CatBoost. For \(T_{\text{eff}}\) task, EstNet34 exhibits minimal fluctuation compared to LightGBM, as evidenced by the length of the lines in the plot. Therefore, EstNet34 not only has higher estimation accuracy than LightGBM but also shows less performance degradation due to anomalous data, indicating stronger robustness. For \(\log\) g task, EstNet66 has smaller errors than CatBoost before adding noise, and after adding noise, the errors of both models are similar. The fluctuation in the evaluation metrics before and after adding noise is slightly larger for EstNet compared to CatBoost. This phenomenon is due to the varying degrees of dispersion in the \(\log\) g and \(T_{\text{eff}}\) labels. The \(\log\) g labels are concentrated with a standard deviation of 0.55, while the \(T_{\text{eff}}\) labels have a large variation with a standard deviation of 10,830. The more concentrated the label values, the more challenging the prediction, which can affect the ability of EstNet to identify anomalous data.

The outstanding performance of EstNet34 on \(T_{\text{eff}}\) is remarkable. In highly noisy and highly dispersed data, EstNet is designed to focus more on non-anomalous data. In contrast, boosting algorithms like LightGBM aim to reduce the fitting error of each base learner \citep{schapire1999brief, mayr2014evolution, duffy2002boosting}, and anomalous samples often cause large fitting errors, leading the base learner to overly focus on these anomalies, which can sometimes be unnecessary.

\section{\textbf{Validation}}
\label{sec:Validation}
Traditional methods for estimating white dwarf parameters primarily rely on theoretical models, which involve fitting the differences between observed and theoretical spectra. These techniques often require high signal-to-noise ratios and high-quality observed spectra. To demonstrate the reliability of EstNet, we obtained spectral data from LAMOST, where white dwarf parameters were estimated using traditional methods \citep{guo2015white, kepler2021white}, and applied EstNet to this data.

\citet{guo2015white} combined three methods, colour–colour cut, LAMOST pipeline classification, and the width of Balmer lines—to select candidate white dwarfs. These white dwarfs were confirmed through visual inspection of spectra. To accurately estimate parameters, spectra with SNR > 10 were selected for absorption line fitting. For DA-type white dwarfs, \(T_{\text{eff}}\) and \(\log\) g were estimated by fitting the line profiles from H\(\beta\) to H\(\epsilon\). In practice, the line profiles of observed and theoretical spectra are normalized using two adjacent points on either side of each absorption line, ensuring that line fitting is unaffected by flux calibration. The atmospheric models used in the fitting process were provided by Koester (2010) \citep{koester2010white}. The widely-used Levenberg-Marquardt nonlinear least-squares method, based on the steepest descent, was employed to fit the line profiles. The best model templates were fitted using the open-source IDL package MPFIT \citep{markwardt2009asp}.

\citet{kepler2021white} utilized SDSS DR16 data to identify and classify white dwarfs and subdwarfs. This dataset includes 2,410 spectra, identifying 1,404 DA-type, 189 DZ-type, 103 DC-type, and 12 DB-type. By simultaneously fitting the photometry and spectra for white dwarfs with SNR \(\geq\) 10 and parallax/error \(\geq\) 4, they estimated effective temperatures and surface gravities. For white dwarfs with M-dwarf companions, the H\(\alpha\) line was excluded from the fitting process due to contamination.

Figure \hyperref[fig:kepler_guo]{\ref*{fig:kepler_guo}} shows the distribution of EstNet predictions compared to parameter estimates obtained using traditional methods. The red solid line represents the identity line. The scatter points of predicted values and true labels are concentrated around the identity line, indicating that the predictions of EstNet are consistent with the parameters estimated using traditional methods. This validation enhances our confidence in EstNet and lays the foundation for large-scale automated parameter estimation in practice.

\section{\textbf{Coclusion}}
\label{sec:Conclusion}
When estimating the parameters of white dwarfs in practice, we find that data not only has low resolution but also an extremely low signal-to-noise ratio. To address this, we designed the EstNet deep learning model, which has three notable advantages.First, we successfully integrated CNN, RNN, and FCN. EstNet captures both the local spatial information and the long-distance dependencies between different absorption lines. This gives it stronger learning capabilities, overcoming the limitations of existing astronomical research that relies solely on models based on CNN and RNN. Second, we designed an adaptive loss mechanism and embedded it into the EstNet model. This allows EstNet to automatically prioritize learning from non-anomalous data and minimize learning from anomalous data.Third, EstNet incorporates Monte-Carlo Dropout, enabling the measurement of uncertainty in output labels. This enhances the interpretability of EstNet and allows users to reasonably assess the reliability of the estimated results.

To demonstrate the performance of EstNet, we conducted reliability analysis, comparative analysis, and robustness analysis. For the \(\log\) g and \(T_{\text{eff}}\) estimation tasks, EstNet outperforms other widely used machine learning algorithms in astronomy across all evaluation metrics. In the \(T_{\text{eff}}\) estimation task, The results of EstNet were less affected by noise compared to other models, highlighting its robustness in handling low-resolution, high-noise, and highly dispersed spectral data. Additionally, the results of EstNet are consistent with the white dwarf parameter estimates obtained using traditional methods by Guo et al. and Kepler et al. \citep{guo2015white, kepler2021white}, further validating the effectiveness of EstNet.

Ongoing survey projects like Large Synoptic Survey Telescope, Sloan Digital Sky Survey-V, and 4-metre Multi-Object Spectrograph Telescope will continue to release more spectral data, a significant portion of which will be low-resolution and low signal-to-noise. In the coming years, the Chinese Space Station Telescope will conduct deep sky surveys, obtaining vast amounts of seamless spectral data with a resolution of 200. Designing more robust parameter estimators to improve automated estimation from massive spectral datasets, especially poor-quality ones, remains our ongoing objective.

\bibliographystyle{aa}
\bibliography{reference.bib}

\end{document}